\documentclass[journal]{IEEEtran}

\ifCLASSINFOpdf
\else
   \usepackage[dvips]{graphicx}
\fi
\usepackage{url}

\usepackage{graphicx}
\usepackage{amsfonts}
\usepackage{amsmath}
\usepackage{bm}
\usepackage{cite}
\usepackage{epstopdf}
\usepackage{flushend}
\usepackage{amssymb}
\usepackage{soul}
\usepackage{color}
\usepackage{threeparttable}
\usepackage{booktabs}
\soulregister\eqref7
\soulregister\cite7

\begin{document}

\title{Joint DOD and DOA Estimation in Slow-Time MIMO Radar via PARAFAC Decomposition}

\author{Feng Xu, \IEEEmembership{Student Member, IEEE}, Sergiy A. Vorobyov, \IEEEmembership{Fellow, IEEE}, Xiaopeng Yang, \IEEEmembership{Senior Member, IEEE}
\thanks{This work was supported in parts by the National Nature Science Foundation of China (Grant No. 61860206012, 61671065 and 31727901), Academy of Finland (Grant: No. 299243), and by the China Scholarship Council. This work was conducted while Feng Xu was a visiting doctoral student with the Department of Signal Processing and Acoustics, Aalto University. 
}
\thanks{Feng Xu and Xiaopeng Yang are with  the School of Information and Electronics, Beijing Institute of Technology, Beijing 100081, China. (e-mail: fengxu@bit.edu.cn;xiaopengyang@bit.edu.cn).}
\thanks{Sergiy A. Vorobyov is with the Department of Signal Processing and Acoustics, Aalto University, Espoo 02150, Finland. (e-mail: svor@ieee.org).}}

\markboth{}
{Shell \MakeLowercase{\textit{et al.}}: Bare Demo of IEEEtran.cls for IEEE Journals}
\maketitle

\begin{abstract}
We develop a new tensor model for slow-time multiple-input multiple output (MIMO) radar and apply it for joint direction-of-departure (DOD) and direction-of-arrival (DOA) estimation. This tensor model aims to exploit the independence of phase modulation matrix and receive array in the received signal for slow-time MIMO radar. Such tensor can be decomposed into two tensors of different ranks, one of which has identical structure to that of the conventional tensor model for MIMO radar, and the other contains all phase modulation values used in the transmit array. We then develop a modification of the alternating least squares algorithm to enable parallel factor decomposition of tensors with extra constants. The Vandermonde structure of the transmit and receive steering matrices (if both arrays are uniform and linear) is then utilized to obtain angle estimates from factor matrices. The multi-linear structure of the received signal is maintained to take advantage of tensor-based angle estimation algorithms, while the shortage of samples in Doppler domain for slow-time MIMO radar is mitigated. As a result, the joint DOD and DOA estimation performance is improved as compared to existing angle estimation techniques for slow-time MIMO radar. Simulation results verify the effectiveness of the proposed method.
\end{abstract}

\begin{IEEEkeywords}
DOD and DOA estimation, factor matrices, PARAFAC, phase modulation matrix, slow-time MIMO radar
\end{IEEEkeywords}

\IEEEpeerreviewmaketitle

\section{Introduction}

\IEEEPARstart{M}{ultiple}-input multiple-output (MIMO) radar \cite{1, 2, 3, 4, 7,6,8,5, 9, 10, 11}, which generally splits into colocated MIMO radar \cite{3} and widely separated MIMO radar \cite{4}, has received a lot of attention over the past decade due to the advantages in multiple targets detection \cite{7}, parameters estimation \cite{6,8} and many other applications \cite{5}. MIMO radar simultaneously emits several orthogonal waveforms via colocated or widely separated antennas to achieve waveform/special diversity. For the case of colocated MIMO radar, the waveform diversity can also be achieved in Doppler domain. The corresponding MIMO radar is named as {\it slow-time MIMO radar} \cite{5, 9, 11}, while the associated waveform design approach is called Doppler division multiple access (DDMA) \cite{10, 11}. The main idea of DDMA is to apply diverse phase modulation values at each transmitter from pulse-to-pulse so that every transmit waveform possesses independent Doppler frequency. Slow-time MIMO radar is approximately equivalent to its conventional MIMO counterpart with reduced Doppler estimation range, but simple waveforme design.

In bistatic colocated MIMO radar, many algorithms for joint direction-of-departure (DOD) and direction-of-arrival (DOA) estimation have been proposed \cite{23,20,24,25,13,14,16,17,18,15}. For example, joint DOD and DOA estimation can be conducted by multiple signal classification (MUSIC) which generally demands two dimensional (2D) spectrum search \cite{23}. By exploiting rotational invariance property (RIP) of signal subspace, estimation of signal parameters via rotational invariance technique (ESPRIT) can be applied to estimate angle information without spectrum search whereas DOD and DOA pairing is still required \cite{20}. In \cite{24}, a generalized algorithm called unitary-ESPRIT (U-ESPRIT) has been introduced to reduce computational complexity. Propagator method (PM) has been also proposed in \cite{25} to avoid singular value decomposition (SVD). The aforementioned algorithms can be regarded as signal subspace-based methods, which normally ignore the multi-linear structure of received data and have poor performance at low signal-to-noise ratio (SNR). A possible solution to overcome these disadvantages is to store the received signals in a tensor. In \cite{13,14,16,17,18,26}, parallel factor (PARAFAC) analysis has been applied to address the problem of poor estimation performance at low SNR. However, conventional tensor model is improper for slow-time MIMO radar because the significantly reduced number of samples in Doppler domain causes a performance loss \cite{26}. Thus, the tensor model-based approach for angle estimation in slow-time MIMO radar needs further investigation.

In this letter, we develop a new tensor model for bistatic slow-time MIMO radar and apply it to joint DOD and DOA estimation. In our model, the received signals are organized in a 3-order tensor. Then modified alternating least squares (ALS) algorithm with extra constant terms is introduced to estimate factor matrices. Finally, the Vandermonde structure of the transmit and receive steering matrices is fully exploited to obtain the angle information from factor matrices. Interestingly, the new tensor model can be regarded as element-wise product of two tensors of different ranks. The first one presumes the same multi-linear structure as that of in the conventional MIMO radar, and the other one contains all phase modulation values for DDMA technique. This enables the proposed method to take advantages of tensor-based algorithms while maintaining the number of samples in Doppler domain. Angle estimation performance for slow-time MIMO radar can hence be significantly improved, which is verified by simulation results.

\section{Slow-Time MIMO Radar Signal Model}

Consider a bistatic MIMO radar with $M$ collocated transmit and $N$ collocated receive antenna elements. Both transmit and receive arrays are uniform linear arrays (ULAs) whose element spacing are half the working wavelength. The steering vectors of the transmit and receive arrays are then denoted by
${\bm{\alpha }}(\varphi ) \triangleq {\left[ {1,{e^{ - j\pi \sin \varphi }}, \cdots, {e^{ - j(M - 1)\pi \sin \varphi }}} \right]^T}$ and ${\bm{\beta }}(\theta ) \triangleq {\left[ {1,{e^{ - j\pi \sin \theta }}, \cdots, {e^{ - j(N - 1)\pi \sin \theta }}} \right]^T}$, where $\varphi$ and $\theta$ are the DOD and DOA, respectively, and ${\left( \cdot  \right)^T}$ denotes the transpose of a matrix/vector. Assuming that there are totally $K$ targets in a range cell of interest, the transmit and receive steering matrices are given, respectively, as ${\bf{A}} \triangleq \left[ {{\bm{\alpha }}({\varphi _1}),{\bm{\alpha }}({\varphi _2}), \cdots ,{\bm{\alpha }}({\varphi _K})} \right]$ and ${\bf{B}} \triangleq \left[ {{\bm{\beta }}({\theta _1}),{\bm{\beta }}({\theta _2}), \cdots ,{\bm{\beta }}({\theta _K})} \right]$.

The matrix of transmit waveforms is denoted by ${\bf{S}}_0 \triangleq {\left[ {{{\bf{s}}_1}, {{\bf{s}}_2}, \cdots, {{\bf{s}}_M}} \right]^T} \in {{\mathbb C}^{M \times L}}$ where $L$ is the number of snapshots per pulse. To achieve waveform diversity in slow-time MIMO radar, a phase modulation matrix ${\bf{W}} \triangleq {\left[ {{{\bf{w}}_1},{{\bf{w}}_2}, \cdots ,{{\bf{w}}_Q}} \right]_{M \times Q}}$ is used at the transmitter during single coherent processing interval (CPI) with $Q$ pulses. The waveform envelopes at all transmit elements are identical. Typically, a linear frequency modulated (LFM) signal ${\bf u} \in {{\mathbb C}^{L \times 1}}$ is used. In $q$-th pulse, $q = 1,2, \cdots ,Q$, the transmitted signal after applying DDMA technique is ${{\bf{S}}_q} = {{\bf w}_q}{\bf u}^T$. According to \cite{5,9,10,11}, the phase modulation matrix from pulse to pulse is 
\begin{equation}
\begin{aligned}&{{\bf{w}}_q} = {e^{j2\pi {\bf{f}}qT}}, \qquad  {\bf{f}} \triangleq [f_1, \cdots, f_M]^T \\
&f_m = \frac{{{f_a}}}{2} \left( - 1 + \frac{{2m - 1}}{M} \right), \; m = 1,2, \cdots, M
\end{aligned}
\end{equation}
where $T$ is the radar pulse duration and $f_a$ is the pulse repetition frequency (PRF). For single scatterer at location $(\varphi_k,\theta_k)$ with Doppler frequency $f_k$ and complex value $\sigma_k^2$ (defined as radar cross section (RCS) fading coefficient), the received signal of $q$-th pulse at the output of $n$-th receive element can be written as
\begin{equation}
{{\bf{x}}_{nq}} = \sigma_k^2{e^{j2\pi {f_k}qT}}{\beta_n}({\theta_k}){{\bm{\alpha }}^T}({\varphi_k}){\bf{S}}_q + {{\bf{n}}_{nq}} \label{6}
\end{equation}
where ${\beta _n}({\theta _k})$ is the $n$-th element of ${\bm \beta}(\theta_k)$ and ${{\bf{n}}_{nq}}$ is the Gaussian white noise. Then the received signal after pulse compression, range gating, and lowpass filtering in slow-time MIMO radar can be expressed as
\begin{equation}
{y_{n{\bar q}m}} = \sigma_k^2{\beta _n}({\theta_k}){{\rm{\alpha }}_m}({\varphi_k}){\bar \gamma_{{\bar q}}}({f_k}) + {\bar z_{n{\bar q}m}}\label{tensor_element}
\end{equation}
where ${{\rm{\alpha }}_m}({\varphi_k})$ and ${\bar \gamma_{{\bar q}}}({f_k})$ are the $m$-th and ${\bar q}$-th elements of the corresponding vectors, ${\bar q} = 1,2, \cdots, {Q \mathord{ \left/{\vphantom {Q M}} \right. \kern-\nulldelimiterspace} M}$, $Q/M$ is assumed to be an integer, ${\bf{\bar \gamma}}({f_k}) \triangleq {\left[ {1,{e^{j2\pi {f_k}T}}, \cdots, {e^{j2\pi ({Q \mathord{\left/{\vphantom {Q M}} \right. \kern-\nulldelimiterspace} M}){f_k}T}}} \right]^T}$, and ${\bar z_{n{\bar q}m}}$ is the noise residue after processing. See Appendix for details.

Therefore, all received signals from $K$ targets can be collected into the following 3-order tensor of size $M \times N \times {Q/M}$:
\begin{equation}
{\cal \bar Y}_s = \sum\limits_{k = 1}^K {\sigma _k^2{ \bm{\alpha }}({\varphi _k}) \circ {\bm \beta} ({\theta _k}) \circ {\bm \bar \gamma} ({f_k})}  + {\cal \bar Z}_s\label{tensor}
\end{equation}
where $\circ $ denotes the outer product and ${\cal \bar Z}_s$ is the noise tensor.

\section{Joint DOD and DOA Estimation for Slow-Time MIMO Radar}

\subsection{Conventional Methods}

Estimators of $\left\{ {{\theta_{k}}} \right\}_{k = 1}^K$ and $\left\{ {{\varphi _{k}}} \right\}_{k = 1}^K$ based on signal subspace algorithms have been conventionally conducted on a per-pulse basis. Using these methods, results can be updated from pulse to pulse. Specifically, we can arrive to the conventional signal model just from the mode-$3$ unfolding (frontal slices) of \eqref{tensor}, given by \cite{16,17}
\begin{equation}
\begin{aligned}
&{\cal \bar Y}_{s(3)} = \left( {{\bf{A}} \odot {\bf{B}}} \right){{\bf{\bar C}}}^T + {{\bf{\bar Z}}_s} \\
&{{\bf{\bar C}}} \triangleq \left[ {{\bf{\bar c}}_1,{\bf{\bar c}}_2, \cdots, {\bf{\bar c}}_{Q/M}} \right]^T \label{matrix}
\end{aligned}
\end{equation}
where ${{\bf{\bar c}}_{{\bar q}}} = {\bf{c}} * {{\bm{\chi }}_{{\bar q}}}$, ${\bf{c}} \triangleq \left[ {\sigma _1^2, \sigma _2^2, \cdots, \sigma _K^2} \right]^T$, ${{\bm{\chi }}_{{\bar q}}} \triangleq \left[ {{e^{i2\pi {f_1}{\bar q}T}}, {e^{i2\pi {f_2}{\bar q}T}}, \cdots, {e^{i2\pi {f_K}{\bar q}T}}} \right]^T$, $*$ stands for the Hadamard product, $\odot$ denotes the Khatri-Rao product (column-wise Kronecker product), and ${{\bf{\bar Z}}_s}$ is the matricized form of ${\cal \bar Z}_s$ of dimension $MN \times Q/M$. Inspecting a single pulse of \eqref{matrix}, e.g., the ${\bar q}$-th pulse, the received signal is
\begin{equation}
{{\bf{\bar y}}_{s{\bar q}}} = \left( {{\bf{A}} \odot {\bf{B}}} \right){{\bf{\bar c}}_{{\bar q}}} + {{\bf{\bar z}}_{s{\bar q}}}\label{matrix_data}
\end{equation}
which coincides with the signal model used in the conventional signal subspace-based angle estimation algorithms.

Let us reshape \eqref{matrix_data} into the following $M \times N$ matrix:
\begin{equation}
{{\bf{\bar Y}}_{s{\bar q}}} = {\bf{B\Sigma }}{{\bf{\Sigma }}_{{\bar q}}}{\bf{A}}^T + {{\bf{\bar Z}}_{s{\bar q}}} \label{11}
\end{equation}
where ${\bf {\Sigma}} \triangleq diag(\bf c)$, ${{\bf{\Sigma }}_{{\bar q}}} \triangleq diag({{\bm{\chi }}_{{\bar q}}})$, and $diag\left( \cdot  \right)$ denotes the operator that builds a diagonal matrix from a column vector. Model \eqref{11} is identical to that in \cite{18} when Doppler effect is added, except for the reduced number of pulses. For signal subspace-based algorithms, this difference has slight influence. However, it may cause serious performance degradation for tensor-based algorithms.

\subsection{Modified Tensor Decomposition-Based Joint DOD and DOA Estimation in Slow-Time MIMO Radar}

To overcome the performance loss caused by the reduced number of pulses, a new tensor model for slow-time MIMO radar is designed. 

Recall \eqref{result}, it can be regarded as an $M$ times downsampling sequence after lowpass filtering in Doppler domain applied in order to avoid Doppler ambiguity. In \cite{333} and \cite{222}, it is shown that PARAFAC decomposition can often be computed by means of a simultaneous matrix decomposition when the tensor is tall in one mode. If the condition $Q\geq MN$ is satisfied in a 3-order tensor ${\cal Y} \in {{\mathbb C}^{M \times N \times Q}}$, the convergence of the ALS algorithm can be improved by applying the singular value decomposition (SVD) of mode-3 matrix ${\bf Y}_{(3)}$. In radar, $Q\geq MN$ is a common case. However, the number of efficient pulses for each transmit and receive channel in slow-time MIMO radar is reduced from $Q$ to $Q/M$ that challenges the condition. In the following, we show that the number of pulses can be restored to $Q$ by interpolation.

Specifically, let the received signal ${{\bf{y}}_{n}}$ for a single receive element in $Q$ pulses be collected. Pulse compression is applied in fast time for each pulse and fast Fourier transform (FFT) is utilized in slow-time.\footnote{An example of the output of these operation will be shown in Fig.~\ref{fig3}.} Phase demodulation and lowpass filtering in Doppler domain are used to separate received signal for each transmit and receive channel. By exploiting $M$ times interpolation column-wisely and applying IFFT, the received signal can be reformulated as ${\sigma_k^2 {\beta _n}({\theta _k}){\alpha _m}({\varphi _k}){e^{j2\pi {f_k}qT}}}$. Then, $\bf W$ is applied again to shift each component with a unique Doppler frequency, which finally gives ${\chi _{nq,m}} ={\sigma _k^2{\beta _n}({\theta _k}){\alpha _m}({\varphi _k}){e^{j2\pi ({\bar f_k}+{f_m})qT}}}$. It is worth noting that the loss of Doppler information with relatively high frequency means the Doppler frequency ${\bar f_k}$ in ${\chi _{nq,m}}$ after interpolation is different with $f_k$ in ${\eta_{nq,m}}$. However, due to the independence of spatial information and Doppler information, this interpolation has no influence on angle estimation.

Therefore, the reformulation of \eqref{tensor_element} for $K$ targets with $M$ samples can be approximately written as
\begin{equation}
{y_{nqm}} \approx  \sum\limits_{k = 1}^K {\sigma _k^2{\beta _n}({\theta _k}){ \gamma _q}({\bar f_k})} \left( {{\alpha _m}({\varphi _k}) w_{m,q}} \right) + { z_{nqm}} \label{newtensor}
\end{equation}
where $w_{m,q}$ is the $(m,q)$-th element of $\bf W$ and ${\bf{ \gamma }}({\bar f_k}) \triangleq {\left[ {1,{e^{j2\pi {\bar f_k}T}}, \cdots, {e^{j2\pi {\bar f_k}(Q-1)T}}} \right]^T}$. Note that there is no index $n$ in $w_{m,q}$, meaning that $w_{m,q}$ is repeated from one receive element to another without any changes. This property will be used further to separate the phase modulation component. Expression \eqref{newtensor} is approximate because of the above mentioned interpolation, which however has no influence on angle estimation due to the independence of spatial and Doppler information.

First, reshape the received signal of the $q$-th pulse in \eqref{newtensor} into matrix form as
\begin{equation}
{{\bf{Y}}_{sq}} = {\bf{B\Sigma }}{{\bf{\Sigma }}_{q}}({{\bf{A}}^T \odot {\bf w}_q^T}) + {{\bf{Z}}_{sq}} \label{Ysq}
\end{equation}
where ${{\bf{\Sigma }}_{q}} \! \triangleq \! diag({{\bm{\chi }}_{q}})$. Note that ${{\bf{A}}^T \odot {\bf w}_q^T} = {\left( {{{\bf{\Omega }}_q}{\bf{A}}} \right)^T}$ where ${{\bf{\Omega }}_q} \triangleq diag({\bf w}_q)$. The $MN \times 1$ vectorized form of \eqref{Ysq} is
\begin{equation}
\begin{aligned}
{{\bf{y}}_{sq}} = \left[ {{{\left( {{{\bf{\Omega }}_q}{\bf{A}}} \right)}} \odot {\bf{B}}} \right]{\bf{c}}_q + {{\bf{z}}_{sq}}  = {{\bf{d}}_q}\left( {{\bf{A}} \odot {\bf{B}}} \right){\bf{c}}_q + {{\bf{z}}_{sq}}
\end{aligned}
\end{equation}
where ${{\bf{d}}_q} \triangleq diag({{\bf{w}}_q} \otimes {{\bf{1}}_{N \times 1}})$, ${{\bf{c}}_{q}} = {\bf{c}} * {{\bm{\chi }}_{q}}$ contains the targets RCS and Doppler information, and ${\bf{1}}_{N \times 1} \in {{\mathbb C}^{N \times 1}}$ is all-one vector.

Let ${\bf{Y}}_s \triangleq [{{\bf{y}}_{s1}}, {{\bf{y}}_{s2}}, \cdots, {{\bf{y}}_{sQ}}]$ denote the received signal for $Q$ pulses. It can be expressed as
\begin{equation}
{\bf{Y}}_s = [({\bf{A}} \odot {\bf{B}}){{\bf{C}}^T}] * {\bf{D}} + {\bf{Z}}_s \label{newmatrix}
\end{equation}
where ${{\bf{C}}} \triangleq \left[ {{\bf{c}}_1, {\bf{c}}_2, \cdots, {\bf{c}}_Q} \right]^T$, ${\bf{Z}}_s \triangleq \left[ {{{{\bf{ z}}}_{s1}}, {{{\bf{z}}}_{s2}}, \cdots, {{{\bf{z}}}_{sQ}}} \right]$, ${\bf{D }} = ({{\bf{I}}_{M}} \odot {{\bf{1}}_{N\times M}}){\bf{W}}$, and  ${{\bf{I}}_{M}}$ denotes $M\times M$ identity matrix. Comparing \eqref{newmatrix} with \eqref{matrix}--\eqref{11}, it is important to stress that the matrix $\bf D$ (from the DDMA technique) is applied on each transmit-receive channel at every pulse. Moreover, the matrices ${\bf{Y}}_s$ and ${\bf{D}}$ in \eqref{newmatrix} can be regarded as the model-3 unfoldings \cite{16,17} of the following two tensors that have different ranks
\begin{equation}
\begin{aligned}
&{\cal Y} = {\cal Y}_{s} * {\cal D}+{\cal Z}_s \\
& {\cal Y}_{s} = {\cal I}_K {\scriptstyle \times
\scriptscriptstyle1}{\bf{A}} {\scriptstyle \times \scriptscriptstyle2}{\bf{B}} {\scriptstyle \times \scriptscriptstyle3}{\bf{C}} \\
& {\cal D}= {\cal I}_M {\scriptstyle \times \scriptscriptstyle1}{\bf{I}}_M {\scriptstyle \times \scriptscriptstyle2}{\bf{1}}_{N\times M} {\scriptstyle \times \scriptscriptstyle3}{\bf{W}}^T
\end{aligned}\label{slowtime2}
\end{equation}
where ${\cal I}_M$ is $M\times M \times M$ identity tensor and symbol ${\times \scriptstyle i}$, stands for the mode-$i$ product of a tensor and a matrix \cite{17}.

The important observation from \eqref{slowtime2} is that the new signal tensor model for slow-time MIMO radar can be regarded as the Hadamard product of two tensors. One of them is identical to the conventional MIMO radar tensor model, while the other one stands for the phase modulation values applied on the transmit elements. It is also worth noting that the mode-2 unfolding (lateral slices) of ${\cal D}$ are identical to $\bf W$, which exactly explains the essence of DDMA technique.

Consequently, $({\theta}_k, {\varphi}_k)$, $k=1, \cdots, K$ can be estimated by fully exploiting the Vandermonde structure of the factor matrices $\bf {A,B}$. Take ${\varphi}_k$ for example. Then two subarrays, one without the last element and the other without the first element of the transmit array, can be formed. Using \eqref{slowtime2}, the received signals for these two subarrays can be expressed as
\begin{equation}
\begin{aligned}
& {\cal Y}_{1} = ({\cal I}_{K} {\scriptstyle \times \scriptscriptstyle1}{\bf{A}}_1 {\scriptstyle \times \scriptscriptstyle2}{\bf{B}} {\scriptstyle \times \scriptscriptstyle3}{\bf{C}}) * {\cal D}_{1} + {\cal Z}_{s,1}\\
& {\cal Y}_{2} = ({\cal I}_K {\scriptstyle \times \scriptscriptstyle1}{\bf{A}}_2 {\scriptstyle \times \scriptscriptstyle2}{\bf{B}} {\scriptstyle \times \scriptscriptstyle3}{\bf{C}}) * {\cal D}_{2}+ {\cal Z}_{s,2}\\
\end{aligned}\label{subarray}
\end{equation}
where ${\cal D}_{1} = {\cal I}_{M-1} {\scriptstyle \times \scriptscriptstyle1}{\bf{I}}_{M-1} {\scriptstyle \times \scriptscriptstyle2}{\bf{1}}_{N\times {(M-1)}} {\scriptstyle \times \scriptscriptstyle3} {\bf W}^T_1$, ${\cal D}_{2} = {\cal I}_{M-1} {\scriptstyle \times \scriptscriptstyle1}{\bf{I}}_{M-1} {\scriptstyle \times \scriptscriptstyle2}{\bf{1}}_{N\times {(M-1)}} {\scriptstyle \times \scriptscriptstyle3} {\bf W}^T_2$ with ${\bf W}_1$ and ${\bf W}_2$ standing for submatrices of $\bf W$ without the last and first row, respectively. Similarly, ${\bf{A}}_1$ and ${\bf{A}}_2$ are submatrices of $\bf A$ without the last and first row, respectively. Since $\bf A$ and $\bf W$ are Vandermonde matrices, ${\bf W}_2 = {{\bf W}_1}{{\bf{\Pi }}_{{W}}} $, ${\bf{A}}_2 = {{\bf{A}}_1}{{\bf{\Gamma }}_{{A}}}$, where ${\bf{\Pi }}_{{W}} \triangleq diag\left( {{{\left[ {{e^{j2\pi \Delta fT}},{e^{j2\pi \Delta f2T}},\cdots,{e^{j2\pi \Delta fQT}}} \right]^T}}} \right)$, $\scriptstyle \Delta f  = f_a/M$, and ${{\bf{\Gamma }}_{{A}} \triangleq diag\left( {\left[ {{e^{-j\pi \sin {\varphi _1}}},{e^{-j\pi \sin {\varphi _2}}},\cdots,{e^{-j\pi \sin {\varphi _K}}}} \right]^T} \right)}$.

Let ${\cal Y}_{A} \triangleq \left[ {\cal Y}_{1} {\scriptstyle \sqcup} {\scriptscriptstyle 1} {\cal Y}_{2} \right]$ where ${\sqcup} {\scriptstyle i}$ stands for the concatenation of two tensors along the $i$-th mode. Particularly, it can be
\begin{equation}
{\cal Y}_{A} = \left[ {\cal I}_K {\scriptstyle \times \scriptscriptstyle1}\left( \begin{array}{l}{{\bf{A}}_1}\\ {{\bf{A}}_2} \end{array} \right) {\scriptstyle \times \scriptscriptstyle2}{\bf{B}} {\scriptstyle \times \scriptscriptstyle3}{\bf{C}} \right] * {\cal D}_{A} + {\cal Z}_{s,A}\label{connect}
\end{equation}
where ${\cal D}_{A} = \left[ {\cal D}_{1} {\scriptstyle \sqcup} {\scriptscriptstyle 3} {\cal D}_{2} \right]$ and ${\cal Z}_{s,A} = \left[ {\cal Z}_{s,1} {\scriptstyle \sqcup} {\scriptscriptstyle 1} {\cal Z}_{s,2} \right]$. Note that ${\cal D}_{A}$ is fixed. Therefore, the $2(M-1) \times N \times Q$ tensor ${\cal Y}_{A}$ can be used to conduct PARAFAC decomposition via the following modified ALS algorithm
\begin{equation}
{\bf{\hat A}}_0 \leftarrow \mathop {\min }\limits_{{\bf{\hat A}}_0} \left\| {{{\cal {Y}}_{A(1)}} - \left[ {({\bf{B}} \odot {\bf{C}}){{{\bf{\hat A}}_0}^T}} \right] * {{\cal{D}}_{A(1)}}} \right\|_F^2{\rm{ }} \label{ALS}
\end{equation}
where ${\bf{A}}_0 \triangleq [{\bf{A}}_1^T, {\bf{A}}_2^T ]^T$, ${\cal {Y}}_{A(1)}$ and ${\cal {D}}_{A(1)}$ are the mode-$1$ unfoldings of the tensors ${\cal {Y}}_{A}$ and ${\cal{D}}_A$, respectively, $\| \cdot \|_F^2$ denoted the Frobenius norm of a matrix, and $\bf\hat A$ stands for an estimate of $\bf A$. When $\bf {B,C}$ are fixed, the objective function in \eqref{ALS} is quadratic in $\bf A$. This property remains while the optimization parameter alternates between $\bf {A,B}$, and $\bf{C}$. Thus, at each alternating step, the objective similar to the one in \eqref{ALS} is quadratic with respect to the optimized matrix parameter, and the corresponding PARAFAC decomposition of ${\cal Y}_{A}$ can be found.

Finally, the $(M-1) \times K$ matrices ${\bf{A}}_1$ and ${\bf{A}}_2$ can be extracted from ${\bf \hat A}_0$ for which the property
\begin{equation}
{\bf{\hat A}}_2 = {{\bf{\hat A}}_1}{{\bf{\Gamma }}_A}\label{RIP}
\end{equation}
should hold. Since ${{\bf{\Gamma }}_A}$ has full rank, the least squares method can be used to estimate it, that is, ${{\bf{\hat \Gamma }}_A} = {\bf{\hat A}}_1^\dag {{{\bf{\hat A}}}_2}$ where ${\left( \cdot  \right)^\dag}$ stands for the pseudo-inverse of a rectangular matrix. Using eigenvalue decomposition of ${\bf{\hat A}}_1^\dag {{{\bf{\hat A}}}_2}$, we find the eigenvalues representing the estimates of the diagonal elements of ${{\bf{\Gamma }}_A}$. These estimates are then used to compute $\left\{ {{\varphi _{k}}} \right\}_{k = 1}^K$, e.g., ${\hat \varphi}_k = j \ln({{\bf{\hat \Gamma }}_A}(k,k))/{\pi}$. Here $\ln(\cdot)$ represents natural logarithm.

The parameters $\left\{ {{\theta _{k}}} \right\}_{k = 1}^K$ can be estimated similarly using matrix $\bf{B}$ instead of $\bf{A}$ in \eqref{subarray}--\eqref{RIP}. Specifically, two subarrays without the last and first elements of the receive array are applied, where the receive steering matrices ${\bf B}_1$, ${\bf B}_2$ are obtained from $\bf B$ by removing the last and first rows, respectively. The 2-th mode concatenation of two tensors from \eqref{slowtime2} for subarrays at the receive side is now reformulated as
\begin{equation}
{\cal Y}_{B} = \left[ {\cal I}_K {\scriptstyle \times \scriptscriptstyle1} {\bf{A}}{\scriptstyle \times \scriptscriptstyle2}\left( \begin{array}{l}{{\bf{B}}_1}\\ {{\bf{B}}_2} \end{array} \right)  {\scriptstyle \times \scriptscriptstyle3}{\bf{C}} \right] * {\cal D}_B + {\cal Z}_{s,B}
\end{equation}
where ${\cal D}_B = {\cal I}_M {\scriptstyle \times \scriptscriptstyle1}{\bf{I}}_M {\scriptstyle \times \scriptscriptstyle2}{\bf{1}}_{{2(N-1)}\times M} {\scriptstyle \times \scriptscriptstyle3}{\bf{W}}^T$ is fixed since it is independent on an index of a receive element, and ${\cal Z}_{s,B}$ is the concatenated noise residue. Using the modified ALS algorithm above, the second factor matrix ${\bf{B}}_0 \triangleq [{\bf{B}}_1^T, {\bf{B}}_2^T ]^T$ can be decomposed. Note that ${\bf B}_2 = {{\bf B}_1}{\bf \Gamma}_B$, where ${\bf \Gamma}_B \triangleq diag\left( {\left[ {{e^{-j\pi \sin {\theta_1}}},{e^{-j\pi \sin {\theta_2}}},\cdots,{e^{-j\pi \sin {\theta _K}}}} \right]^T} \right)$. Then the diagonal elements of ${\bf \Gamma}_B$ are estimated by computing the eigenvalues of matrix ${\bf{\hat B}}_1^\dag {{{\bf{\hat B}}}_2}$, and $\left\{ {{\theta _{k}}} \right\}_{k = 1}^K$ are estimated as ${\hat \theta}_k = j\ln({{\bf{\hat \Gamma }}_B}(k,k))/{\pi}$.

Finally, the application of the ALS algorithm above requires the uniqueness\cite{16,17} of PARAFAC decomposition. A weak upper bound on its maximum rank $K$ is given as
\begin{equation}
\min(M,K)+\min(N,K)+\min(Q,K)\geq 2K+2
\end{equation}
which can also be written as $K \leq \min \{MN,MQ,NQ\}$. If $Q\geq MN$, which is a common case in radar, the maximum number of targets that can be resolved is almost surely $K = MN$.

\subsection{Computational Complexity Analysis}
The proposed algorithm for joint DOD and DOA estimation in slow-time MIMO radar requires PARAFAC decomposition and ESPRIT-aided method to compute phase rotations of targets. During each iteration of the ALS algorithm, the number of flops is ${\cal O}(KMNQ)$ \cite{111}. The number of flops for matrix computation, i.e., for computing ${{\bf{\Gamma }}_A}$ and ${{\bf{\Gamma }}_B}$ is ${\cal O}(K^3)$. In total, the number of flops needed in our algorithm is ${\cal O}(KMNQX+K^3)$, where $X$ is the number of iterations. Note that the proposed algorithm requires to perform ALS two times, and it is useful to apply the modified ALS algorithm with better convergence. We refer to \cite{13,16} and the references therein for more details. Table~\ref{tab1} summarizes the computational complexity of the proposed algorithm as well as the complexity of most the existing MIMO techniques.

\begin{table}
\centering
\setlength{\tabcolsep}{8mm}
\begin{threeparttable}
\caption{Computational Complexity of the Existing Algorithms} \label{tab1}
\centering
\begin{tabular}{cc}
\toprule[1pt]
\ Method & Complexity \\
\hline
MUSIC of \cite{23}\tnote{+} & ${\cal O}((MN)^3+M^2N^2QZ)$\\
ESPRIT of \cite{20} & ${\cal O}(M^2N^2(Q+L))$ \\
U-ESPRIT of \cite{24} & ${\cal O}((M^2N^2(Q+2L))$  \\
PM of \cite{25} & ${\cal O}(M^2N^2(K+L)+K^3)$ \\
PARAFAC\tnote{*} of \cite{26} & ${\cal O}(KMNQX)$ \\
Proposed\tnote{*} & ${\cal O}(KMNQX+K^3)$ \\
\bottomrule[1pt]
\end{tabular}
 \begin{tablenotes}
        \footnotesize
        \item[+] $Z$ is the number of grids for spectrum search
        \item[*] $X$ is the number of iteration in ALS algorithm
      \end{tablenotes}
    \end{threeparttable}
\end{table}

\section{Simulation Results}
\begin{figure}
\centerline{\includegraphics[width=\columnwidth]{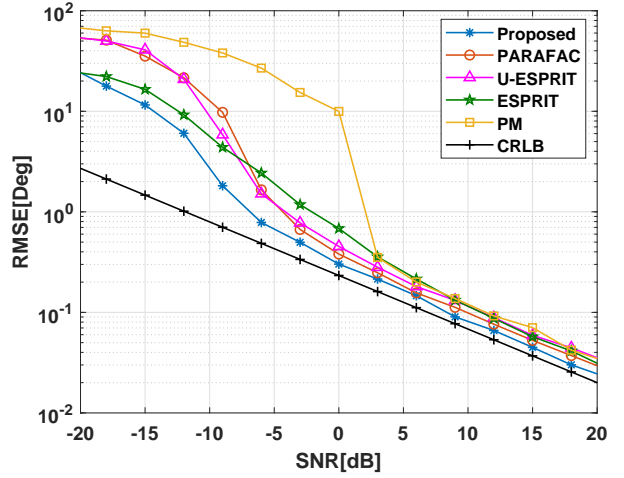}}
\caption{RMSE versus SNR from -20dB to 20dB, 200 trials.}
\label{fig1}
\end{figure}
\begin{figure}
\centerline{\includegraphics[width=\columnwidth]{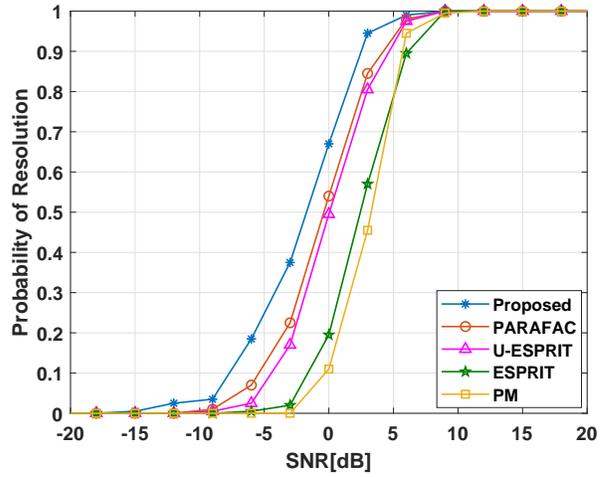}}
\caption{Probability of resolution versus SNR from -20dB to 20dB, 200 trials.}
\label{fig2}
\end{figure}

We demonstrate here the angle estimation performance of the proposed method in comparison to conventional algorithms including PM \cite{25}, ESPRIT \cite{20}, U-ESPRIT\cite{24}, and PARAFAC\cite{18}. The Cramer-Rao lower bound (CRLB) for bistatic MIMO radar is also provided  \cite{27}. Throughout our simulations, a slow-time MIMO radar with $M = 8, N = 10$ antenna elements is considered. A chirp signal with $B = 40$~MHz and $T = 10$~us is used as a waveform envelope $\bf u$. In each CPI, there are $Q=80$ pulses with ${f_a} = 50$~KHz. We assume $K=2$ targets that follow  Swerling~I model \cite{19}, and thus, $\sigma_k^2$ is chosen from a Gaussian distribution as a complex value, which remains fixed from pulse to pulse. The normalized Doppler frequencies of the targets are $f_k = [0.02,-0.05]$. The number of Monte Carlo trials is $P = 200$.

In our first example, the DODs and DOAs of the targets are ${\varphi}_k = [-30^{\circ}, 25^{\circ}]$ and ${\theta}_k = [-15^{\circ}, 20^{\circ}]$, respectively. The root mean square errors (RMSEs) of ${\varphi}_k$ and ${\theta}_k$ are computed separately and then combined.
As can be seen in Fig.~\ref{fig1}, where RMSE is shown versus SNR, the proposed method achieves better performance, especially, at low SNR. Thus, higher estimation accuracy can be achieved by the proposed method. The PM, ESPRIT, and U-ESPRIT methods can be regarded as generalized signal subspace-based approaches, which are sensitive to low SNR. The PARAFAC method exploits the multilinear structure of the received data and avoids degradation at low SNR, but the number of pulses in this method is reduced. By our method, we recover the number of pulses by $M$ times. 

In our second example, the probability of resolution of two closely spaced targets with ${\varphi}_k = [20^{\circ}, 21^{\circ}]$ and ${\theta}_k = [15^{\circ}, 16^{\circ}]$ is analyzed. Two targets are considered to be resolved when $\left\| {{{\hat \theta }_k} - {\theta _k}} \right\| \le \left\| {{\theta _1} - {\theta _2}} \right\|/2$ and $\left\| {{{\hat \varphi }_k} - {\varphi _k}} \right\| \le \left\| {{\varphi _1} - {\varphi _2}} \right\|/2, \, k=1,2$. It can be seen from Fig.~\ref{fig2} that all methods exhibit resolution with probability 1 at high SNR, but the proposed method surpasses other methods as it has the lowest SNR threshold. The improved ability of resolving two closely spaced targets can be regarded as the advantage resulted from combining PARAFAC and ESPRIT. Moreover, the concatenation of tensors in \eqref{connect} for different subarrays approximately doubles the number of samples. Thus, the proposed method achieves higher accuracy and better resolution for angle estimation.

\section{Conclusion}
A new tensor model for slow-time MIMO radar that enables improved joint DOD and DOA estimation for multiple targets has been proposed. This tensor can be regarded as an element-wise product of two tensors, where only one of them contains the angular parameters of interest. The model enables us to use PARAFAC decomposition with ESPRIT, and to address the problem of shortage of samples in Doppler domain for slow-time MIMO radar. As a result, the angle estimation performance has been improved as compared to the existing techniques.

\section{Appendix}
\begin{figure}
\centerline{\includegraphics[width=\columnwidth]{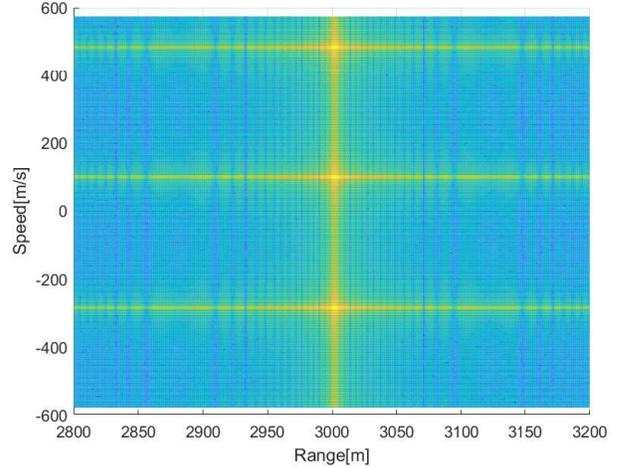}}
\caption{Range-Doppler map for single receive element and $M = 3$ transmit elements in S-band slow-time MIMO radar with single target at $R = 3000m$ moving with velocity $v = 100m/s$. An LFM signal with bandwidth $B = 40MHz$ and $T = 1.6us$ is applied. $Q = 150$. Three peaks at same range cell are generated with different Doppler frequencies determined by $f_m$.}
\label{fig3}
\end{figure}
\begin{figure}
\centerline{\includegraphics[width=\columnwidth]{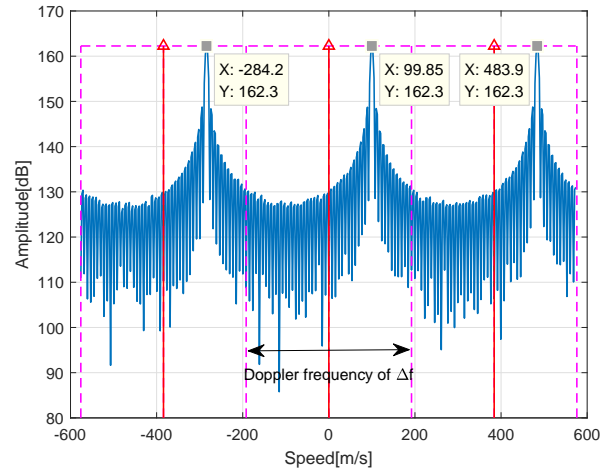}}
\caption{Doppler slice at the range cell of interest, each peak in blue line denotes a single transmit channel. Pink lines define the Doppler space distributed to each transmit channel. Red lines give the Doppler frequency $f_m$ shifted by the phase modulation matrix (should be zero Doppler in their own range-Doppler map).}
\label{fig4}
\end{figure}

To obtain \eqref{tensor_element}, rewrite \eqref{6} as
\begin{equation}
\begin{aligned}
{{\bf{x}}_{nq}} &= \sigma _k^2{e^{j2\pi {f_k}qT}}{\beta _n}({\theta _k}){{\bm{\alpha }}^T}({\varphi _k}){{\bf w}_q}{\bf{u}}^T + {{\bf{n}}_{nq}} \\
& =  \left( {\sigma _k^2{\beta _n}({\theta _k})\sum\limits_{m = 1}^M {{\alpha _m}({\varphi _k}){e^{j2\pi ({f_k} + {f_m})qT}}} } \right){{\bf{u}}^T} + {{\bf{n}}_{nq}} . \label{appendix1}
\end{aligned}
\end{equation}

Applying matched-filtering to ${{\bf{x}}_{nq}}$ in fast time, we have
\begin{equation}
\begin{aligned}
{{\bf{\bar x}}_{nq}}(t) &= {\eta _{nq}}\int_{ - \infty }^\infty  {{\bf{u}}^T(t'){\bf{u}}^T(t - t')}dt'  + {{\bf{\bar n}}_{nq}}(t)\\
& = {\eta _{nq}}{{\bf F}(t)}+{{\bf{\bar n}}_{nq}}(t)
\end{aligned}
\end{equation}
where ${\eta _{nq}} \triangleq {\sigma _k^2{\beta _n}({\theta _k})\sum\limits_{m = 1}^M {{\alpha _m}({\varphi _k}){e^{j2\pi ({f_k} + {f_m})qT}}} }$ and ${{\bf{\bar n}}_{nq}}(t)$ is noise residue after matched-filtering. Since $\bf u$ is an LFM signal, the integral term ${{\bf F}(t)}  = \int_{ - \infty }^\infty  {{\bf{u}}^T(t'){\bf{u}}^T(t - t')}dt'$ is known to be approximately a $sinc$ function, and its peak indicates the range cell of the target.

Then, the concatenation of ${{\bf{\bar x}}_{nq}}$ in a single CPI with $Q$ pulses forms a $Q \times L$ matrix, i.e., 
$[{{\bf{\bar x}}^T_{n1}},{{\bf{\bar x}}^T_{n2}},\cdots,{{\bf{\bar x}}^T_{nQ}}]^T$. If fast Fourier transform (FFT) is applied to this matrix column-wisely, $M$ peaks with different Doppler frequencies can be found at the slice of target range cell. Each of the peaks corresponds to a unique transmit element as shown in the range-Doppler map shown in Fig.~\ref{fig3}. The distance between two adjacent peaks in Doppler domain is determined by $\Delta f = f_a/M$. Owing to this Doppler frequency shifts for different transmitted waveforms, it is possible to distinguish each transmit channel at the receiver via filtering in Doppler domain.

For any $m$-th transmit element, the phase demodulation in Doppler domain is ${{\bf{\bar x}}_{nq,m}} = {{\bf{\bar x}}_{nq}}{e^{ - j2\pi {f_m}qT}}$. Equivalently
\begin{equation}
{{\bf{\bar x}}_{nq,m}}(t)  ={\eta _{nq,m}}{\bf{F}}{{(t) + }}{\kappa _{nq,m}}{\bf{F}}{{(t)}} + {{\bf{\bar z}}_{nq,m}}(t)\label{fa}
\end{equation}
where ${\eta _{nq,m}} \triangleq {\sigma _k^2{\beta _n}({\theta _k}){\alpha _m}({\varphi _k}){e^{j2\pi {f_k}qT}}}$, ${\kappa _{nq,m}} \triangleq \sigma _k^2{\beta _n}({\theta _k})\sum\limits_{m' \ne m} {{{\rm{\alpha }}_{m'}}({\varphi _k}){e^{j2\pi ({f_{m'}} - {f_m} + {f_k})qT}}}$, and ${{\bf{\bar z}}_{nq,m}}(t)$ is the noise residue. After demodulation, each of $M$ transmit channels is shifted to baseband in Doppler domain (with a frequency of $f_k$). A reduced efficient Doppler range of $\left[ { - \frac{{{f_a}}}{{2M}},\frac{{{f_a}}}{{2M}}} \right]$ is distributed to every transmit element (see Fig.~\ref{fig4} for more details). By applying lowpass filtering with cutoff frequency $\Delta f$, the second term ${\kappa _{nq,m}}{\bf{F}}{\rm{(t)}}$ in \eqref{fa} can be omitted. Hence, the received signal from $m$-th transmit element to $n$-th receive element at $q$-th pulse can be expressed as
\begin{equation}
{{\bf{\bar y}}_{nq,m}}(t) = {\eta_{nq,m}}{\bf{F}}{{(t)}} + {{\bf{\bar z}}_{nq,m}}(t)
\end{equation}

By range gating, the received signal is further expressed as
\begin{equation}
\bar y_{nq,m} = {\eta _{nq,m}} + \bar z_{nq,m}
\end{equation}
where $\bar z_{nq,m}$ is the slice of ${{\bf{\bar z}}_{nq,m}}(t)$ at the target range cell. Note that in order to avoid ambiguous Doppler returns, it is necessary to ensure that the highest Doppler frequency of interest $f_k$ is smaller than $\Delta f/2$. This implies that the DDMA technique achieves waveform diversity for slow-time MIMO radar at the cost of reduced Doppler frequency estimation range.

Another disadvantage is the decrease of the number of samples in Doppler domain. Recall ${\eta _{nq,m}}$, we have ${\eta _{nq,m}} = \sigma _k^2{\beta _n}({\theta _k}){\alpha _m}({\varphi _k}){e^{j2\pi \frac{{{f_k}}}{{{f_a}}}q}}, \; q = 1,2,\cdots,Q$. Clearly, this is a discrete signal with sampling rate $f_a$. The uniformly divided Doppler space then leads to the decline of sampling rate to $f_a/M$, i.e., ${{\eta }_{nq,m}} = \sigma _k^2{\beta _n}({\theta _k}){\alpha _m}({\varphi _k}){e^{j2\pi \left( {\frac{{{f_k}}}{{{f_a}/M}}} \right)\frac{q}{M}}}$. Therefore,
\begin{equation}
{{\bar \eta }_{n\bar q,m}} = \sigma _k^2{\beta _n}({\theta _k}){\alpha _m}({\varphi _k}){e^{j2\pi \bar q{f_k}T}},\bar \; q = 1,2,\cdots,Q/M\label{result}
\end{equation}
where the number of efficient pulses is reduced to only $Q/M$. Here $Q/M$ is assumed to be an integer. Considering the noise term, the result in \eqref{tensor_element} is obtained.

\bibliographystyle{IEEEtran}
\bibliography{IEEEabrv,Ref}

\end{document}